\documentclass[useAMS,fleqn,usenatbib]{mnras}
\usepackage{times}
\usepackage{amsmath}
\usepackage{bm}
\usepackage{url}
\usepackage{graphicx}
\usepackage{hyperref}

\input amssym.tex

\mathchardef\mhyphen="2D

\newcommand{\origami}{{\scshape origami}}
\newcommand{\bq}{\textbf{q}}

\newcommand{\bx}{\textbf{x}}

\newcommand{\hmpc}{\,$h^{-1}$\,Mpc}

\newcommand{\camb}{{\scshape camb}}

\newcommand{\bnabla}{\mbox{\boldmath $\nabla$}}
\newcommand{\bPsi}{\mbox{\boldmath $\Psi$}}

\newcommand{\divqpsi}{{\bnabla_q\cdot{\bPsi}}}

\newcommand{\dlin}{\delta_{\rm lin}}

\newcommand{\psiz}{\psi_{\rm Z}}
\newcommand{\psitwo}{\psi_{\rm 2LPT}}

\newcommand{\psisc}{\psi_{\rm sc}}
\newcommand{\psimusc}{\psi_{\rm musc}}

\newcommand{\rip}{R_{\rm ip}}

\newcommand{\alpt}{{\scshape alpt}}
\newcommand{\SC}{{\scshape sc}}
\newcommand{\muscle}{{\scshape muscle}}
\newcommand{\cosmopy}{{\scshape CosmoPy}}

\chardef\til=`\~

\begin{document}

\title[Multiscale spherical collapse realizations]{Truthing the stretch: Non-perturbative cosmological realizations with multiscale spherical collapse}

\author[Mark C.\ Neyrinck]
{Mark C.\ Neyrinck$^1$\\
$^{1}$Department of Physics and Astronomy, The Johns Hopkins University, Baltimore, MD 21218, USA}


\maketitle

\begin{abstract}
Here we present a simple, parameter-free, non-perturbative algorithm that gives low-redshift cosmological particle realizations accurate to few-Megaparsec scales, called \muscle\ (MUltiscale Spherical ColLapse Evolution). It has virtually the same cost as producing $N$-body-simulation initial conditions, since it works with the `stretch' parameter $\psi$, the Lagrangian divergence of the displacement field.
 It promises to be useful in quickly producing mock catalogs, and to simplify computationally intensive reconstructions of galaxy surveys.  \muscle\ applies a spherical-collapse prescription on multiple Gaussian-smoothed scales. It achieves higher accuracy than perturbative schemes (Zel'dovich and 2LPT), and, by including the void-in-cloud process (voids in large-scale collapsing regions), solves problems with a single-scale spherical-collapse scheme.  Slight further improvement is possible by mixing in the 2LPT estimate on large scales. Additionally, we show the behavior of $\psi$ for different morphologies (voids, walls, filaments, and haloes). A Python code to produce these realizations is available at \url{http://skysrv.pha.jhu.edu/~neyrinck/muscle.html}. 
\end{abstract}

\begin {keywords}
  large-scale structure of Universe -- cosmology: theory
\end {keywords}

\section{Introduction}
In the current paradigm, structures on extragalactic scales arise from the stretching and bunching-together of the `dark-matter sheet.' A conceptual understanding of this process is a fundamental goal of astronomy, and is closely tied to several other topics in cosmology and galaxy formation. A Lagrangian picture, following particles (i.e.\ mesh locations on this `dark-matter sheet') \citep{ShandarinEtal2011, AbelEtal2012} instead of fixed comoving positions, is essential to understanding this process. Approximations to full $N$-body dynamics are useful both for this understanding, and to produce approximate $N$-body realizations, e.g.\ for mock galaxy catalogs \citep[e.g.][]{KitauraEtal2014patchy}, and for fast explorations of large parameter spaces for Bayesian inference of the initial-conditions (IC) density field \citep[e.g.][]{KitauraEnsslin2008,KitauraEtal2012,Kitaura2013,HessEtal2013,JascheWandelt2013,LeclercqEtal2015}.

The Zel'dovich approximation \citep[][ZA]{Zeldovich1970}, linear-order LPT (Lagrangian perturbation theory) already captures much of the essential physics of structure formation \citep[e.g.][]{White2014}, producing a cosmic web accurately to rather small scales. Improvements on ZA in various regimes have been proposed. Going to second order (2LPT) improves accuracy for small fluctuations, i.e.\ at large scales and early times, useful for IC generation \citep{Scoccimarro1998,CrocceEtal2006}. At small scales, truncating the power spectrum shortward of the nonlinear scale \citep{KofmanEtal1992}, or using the adhesion model \citep{KofmanShandarin1988,KofmanEtal1992,Shandarin2009,ValageasBernardeau2011,HiddingEtal2015} suppresses unphysical overcrossing of particles in collapsed structures. 

Many of these approaches work with the Lagrangian divergence of the displacement field, $\psi(\bq)\equiv\divqpsi(\bq)$. Here, the displacement field is $\bPsi(\bq) = \bx(\bq)-\bq$, where $\bq$ denotes Lagrangian (IC) coordinates of a mass element, and $\bx$ denotes its Eulerian, final coordinates. A recent approach \citep[][N13]{Neyrinck2013} works non-perturbatively with $\psi$, using a low-density limit of the spherical-collapse (\SC) evolution of a mass element, found by \citet{Bernardeau1994}. This \SC\ relationship was also investigated by \citet{ProtogerosScherrer1997} and in the context of IC reconstructions by \citet{MohayaeeEtal2006}.

Importantly, displacement-divergence ($\psi$-based) schemes giving cosmological realizations are essentially as fast as producing the initial conditions for an $N$-body simulation.  They have three steps: (1) Generate a pixelated linear-theory density field $\dlin$ at the desired redshift, consistent with a linear power spectrum. (2) Estimate $\psi$ from $\dlin$. (3) Generate the final displacement field $\bPsi$ with an inverse-divergence operator, e.g.\ using an FFT; apply $\bPsi$ to particles on a regular lattice to get their final positions.

In the ZA, $\psiz(\dlin)=-\dlin$; in 2LPT, there is a non-local functional giving $\psi$, roughly parabolic in the local $\dlin$ (N13). In the \SC\ prescription, for each particle (IC density field pixel),
\begin{align}
\psisc(\dlin)& = 3\sqrt{1-(2/3)\dlin}-3, \dlin < 3/2;\nonumber \\
 		 & = -3,\ \dlin \ge 3/2.
\label{eqn:sc}
\end{align}
Or, (over)compactly,
\begin{equation}
\psisc(\dlin)\equiv3~{\rm Re}\sqrt{1-(2/3)\dlin}-3.
\label{eqn:screal}
\end{equation}
Setting $\psisc=-3$ produces a collapsed volume element as closely as possible when dealing only with $\psi$, since $\psi=-3$ in three dimensions if adjacent particles coincide.

\SC\ successfully reins in particles, preventing overcrossing on the interparticle scale, even when $\dlin$ is non-perturbatively large. It also gets densities remarkably right in voids, unlike ZA (overevacuating them), and 2LPT (underevacuating them) \citep{SahniShandarin1996}. When applied far enough outside the perturbative regime of small fluctuations, 2LPT even produces overdensities in void centers (N13). However, when applied at the single, highest Lagrangian resolution of a nonlinear-resolution density field, \SC\ has problems as well. It gives realizations with low power on large scales; also, the Fourier-space cross-correlation coefficient departs from unity at larger scales than in ZA.

Several other techniques have been proposed recently to produce approximate, fast $N$-body or halo density fields, such as {\scshape pinocchio} \citep{MonacoEtal2002, MonacoEtal2013}, {\scshape cola} \citep{TassevEtal2013}, and {\scshape qpm} \citep{WhiteEtal2014}. \citet[][KH13]{KitauraHess2013} found one insightful fix to the \SC's issues, calling it Augmented Lagrangian Perturbation Theory (\alpt). In Fourier space,
\begin{equation}
 \psi_{\rm ALPT}(k)=G(k)\psitwo(k) + [1-G(k)]\psisc(k),
 \label{eqn:psialpt}
\end{equation}
an interpolation in scale between 2LPT, which excels on large scales with small fluctuations, and \SC, which gets small scales right. Here $G(k)$ is a Gaussian smoothing kernel. \alpt\ is a key ingredient in the {\scshape patchy} algorithm \citep{KitauraEtal2014patchy}, which produces mock galaxy catalogs using some other novel, useful prescriptions to treat redshift-space distortions and galaxy biasing.

\section{Method and Results}
Here, we give an alternative fix to the \SC\ prescription: making it multiscale. As we show below, the trouble with the \SC\ prescription is that it is applied at the single scale of the interparticle spacing, preventing voids in clouds \citep[voids within larger-scale collapsing regions, e.g.][]{ShethvandeWeygaert2003} from properly collapsing.

MUltiscale Spherical-ColLapse Evolution (\muscle) additionally checks scales larger than the interparticle scale for collapses, thus including the void-in-cloud process.  Like the above schemes, \muscle\ works with step (2), the estimation of $\psi(\dlin)$, with $\dlin$ defined on an interparticle separation (resolution of the initial density field), $\rip$. The prescription for $\psi$ is
\begin{align}
\psimusc(\dlin) &= 3\left(\sqrt{1-\frac{2}{3}\dlin}-1\right), & \dlin < 3/2\ {\rm and}\ \nonumber \\
	& & G_r(\dlin) < 3/2, \forall r\ge \rip; \nonumber \\
 			         & = -3,~  {\rm otherwise}.
\label{eqn:sccond}
\end{align}
$G_r(\dlin)$ denotes the $\dlin$ field, Gaussian-filtered at scale $r$. Note that generally, $\psimusc$ will have nonzero mean, so an additional step is to subtract a constant from all $\psimusc>-3$ to give zero mean.

In practice, a finite, and preferably small, number of $r>\rip$ have to be tried. As implemented, we search for collapses at  $r=2^n\rip$, for integers $n\ge0$. The Gaussian smoothing is done in Fourier space, requiring an FFT and inverse FFT.  
We also tried using a cubic top-hat filter, averaging together sets of 8 particles on larger scales. This did visually localize halo regions better than the Gaussian smoothing, but the below cross-correlation performed more poorly.

A large factor between smoothing scales searched for collapses gives greater speed, while a small factor captures more collapses, giving more accuracy. At our fiducial factor of 2, generating a random $\Lambda$CDM 256\hmpc\ realization with $256^3$ particles at $z=0$ with the \muscle\ Python code searched 7 different smoothing scales; after no collapses were found at 64\hmpc, no larger scales were necessary. On one 2.6 GHz processor, generating this realization with ZA, 2LPT, and \muscle\ took 14, 28, and 59 seconds, respectively.

\begin{figure}
    \begin{center}
    \includegraphics[width=\columnwidth]{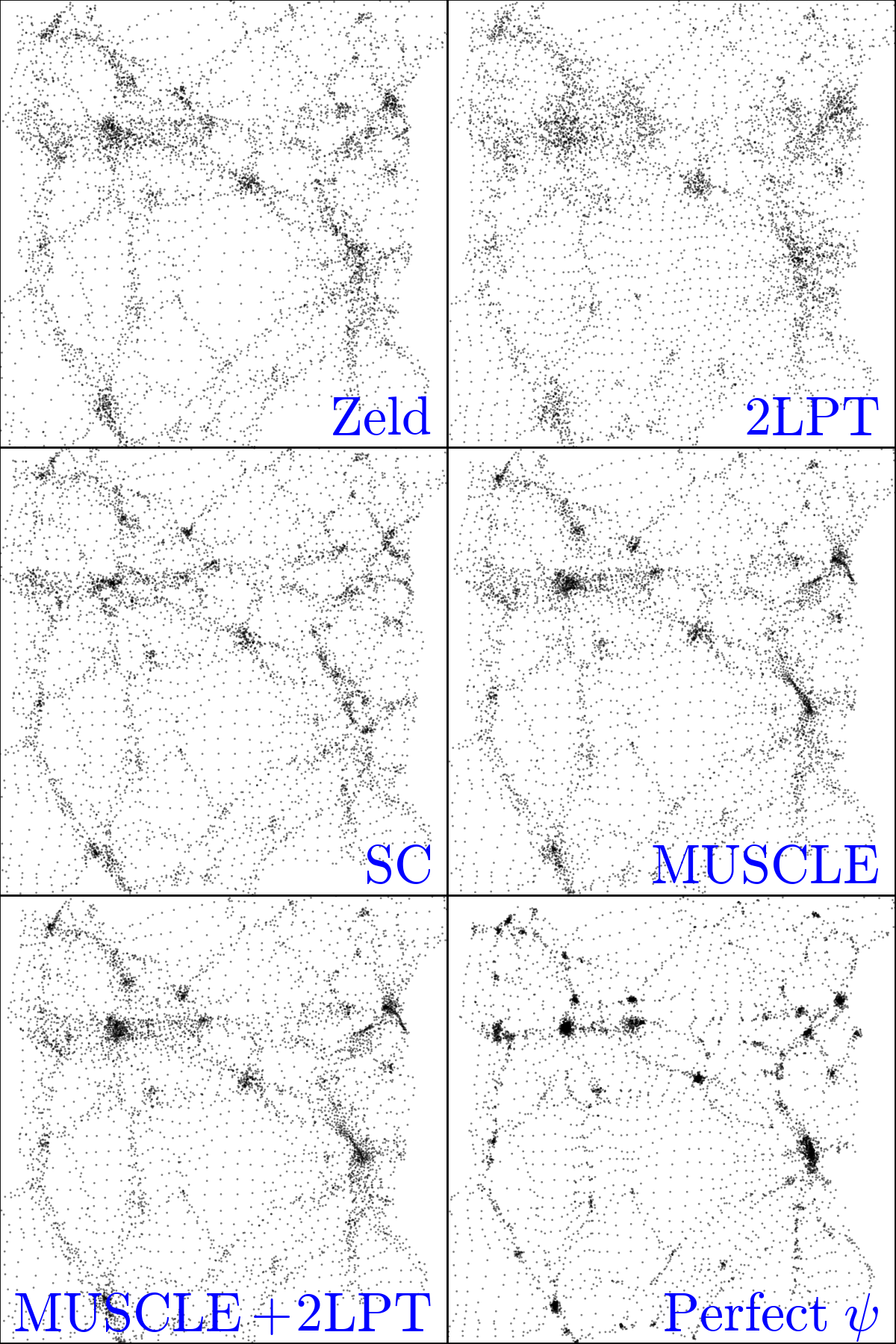}
    \includegraphics[width=0.5\columnwidth]{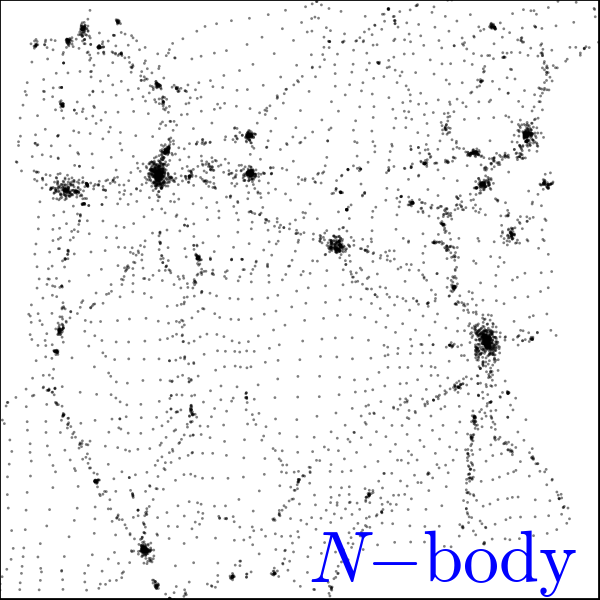}
  \end{center}  
  \caption{Initially square 2D patches, 60\hmpc\ on a side, of 256$^3$-particle realizations made using approximate schemes using the displacement-divergence $\psi$. In `\muscle +2LPT,' $\psi_{\rm MUSCLE}$ is interpolated in scale as in \alpt (shown here as well, using \SC\ instead of \muscle), both with smoothing lengths of 30\hmpc. `Perfect $\psi$' shows positions after removing the curl of the displacement field, and shows  `$N$-body' shows the actual simulation.}
  \label{fig:musclepos}
\end{figure}

\begin{figure}
    \begin{center}
    \includegraphics[width=\columnwidth]{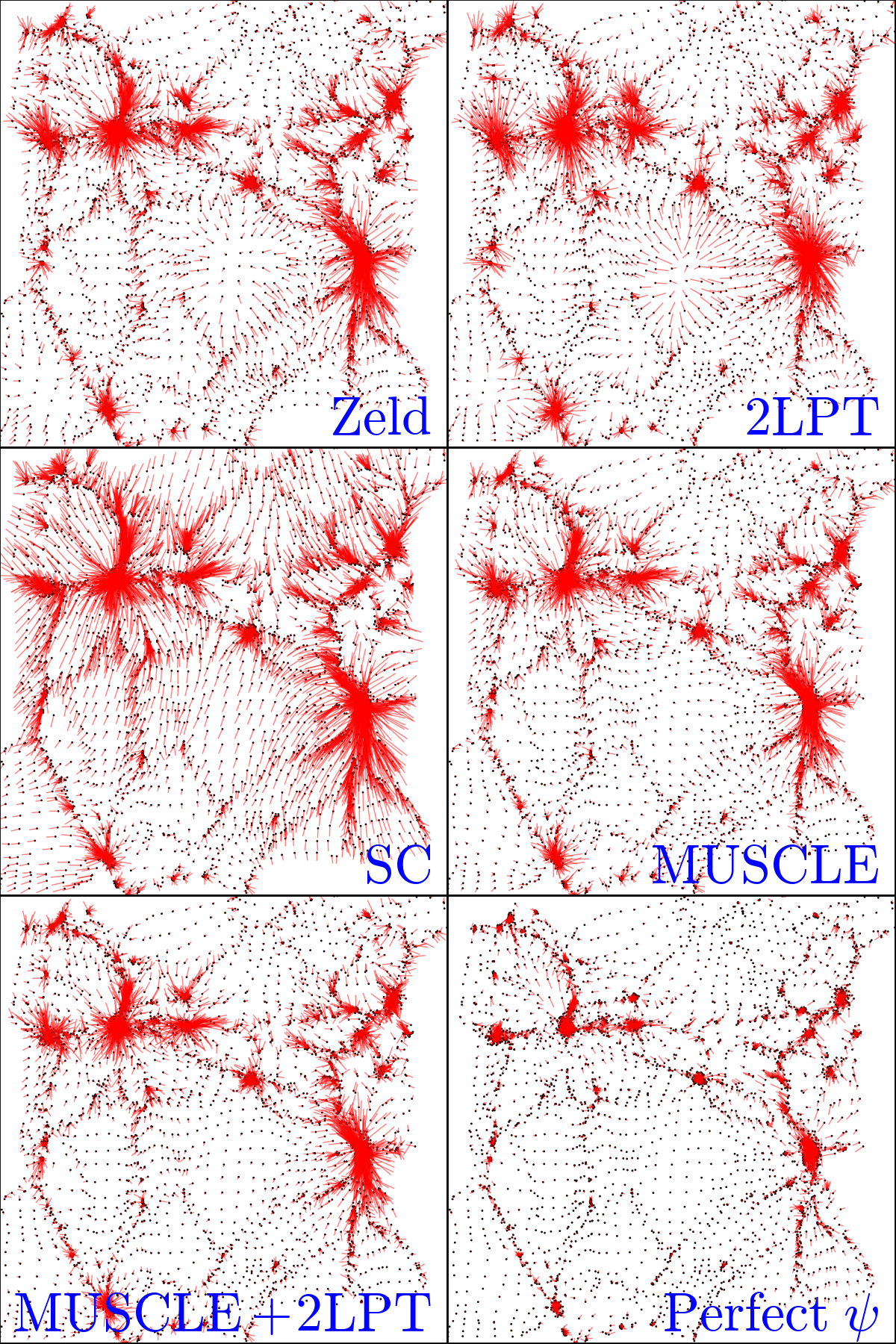}
  \end{center}  
  \caption{Same as Fig.\ \ref{fig:musclepos}, except with a red line joining the position of each particle in the full $N$-body result to the approximation.}
  \label{fig:muscledisp}
\end{figure}

Fig.\ \ref{fig:musclepos} shows 2D patches of 3D particle realizations at redshift $z=0$ using various approximate schemes from the initial conditions. It also shows the full $N$-body results, and `Perfect $\psi$' -- this is generated by measuring the actual $\psi$ field from each particle's $N$-body position using an FFT, and inserting it into step 2 above. This procedure zeroes the curl of the displacement field, but preserves all information in the displacement-divergence $\psi$. See \citet{Chan2014} for another investigation of zeroing the curl. In Fig.\ \ref{fig:muscledisp}, a red line is drawn from each particle's actual $N$-body final position to that in the approximate scheme.

In voids, densities (visible from particle separations) are too low in the ZA (Zeld), and too high in 2LPT. Haloes are puffy in Zeld, and even puffier in 2LPT. In this simulation, the interparticle scale is about 0.8\hmpc, with $256^3$ particles in a 200\hmpc\ box. The linear-theory variance on this scale much exceeds unity, $\sigma_{\rm lin}^2=7.3$; it is not surprising that a higher-order perturbative scheme (2LPT) fails worse than a lower-order scheme (Zeld) when the perturbative parameter is large. \SC, as implemented in Eq.\ (\ref{eqn:sc}), gets densities right in voids, and pulls in overcrossed particle positions in haloes, but as Fig.\ \ref{fig:muscledisp} shows, rather large-scale displacements are wrong in \SC. In \muscle, the large-scale displacement problems are largely solved. This shows that the major deficiency in \SC\ is in its treatment of the void-in-cloud process.

We also show results of `\muscle+2LPT', using the \alpt\ strategy of interpolating $\psi$ between $\psi_{\rm 2LPT}$ on large scales and $\psi_{\rm MUSCLE}$ on small scales, as in Eq.\ \ref{eqn:psialpt}. The best interpolation smoothing parameter using \muscle\ at $z=0$ seems to be about 30\hmpc, judging by $R(k)$ as shown below in Fig.\ \ref{fig:crosscorr}. We also observed gains in $R(k)$ in \alpt\ (using \SC\ instead of \muscle) on small scales up to this smoothing length, although KH13 found that a smaller smoothing was optimal. So, we use the same smoothing length for each. Interpolating \muscle\ with 2LPT gives a few-percent boost in $R(k)$ on non-linear scales, and gives perhaps a slight visual improvement over \muscle, in Figs.\ \ref{fig:musclepos} and \ref{fig:muscledisp}. A smaller interpolation scale, $\lesssim$10\hmpc, resulted in 2LPT visibly corrupting large haloes and voids.

In Fig.\ \ref{fig:crosscorr}, we show measures of the agreement between the $N$-body result and various approximations. At top, we show the Fourier-space cross correlation, $R(k) = P_{\delta\times\delta^\prime}/\sqrt{P_\delta P_{\delta^\prime}}$ between the $z=0$ density field and the various approximate density fields, using nearest-gridpoint density assignment on a 128$^3$ grid. $P_{\delta\times\delta^\prime}$ here is the cross-power between $\delta$ and $\delta^\prime$. Judging by $R(k)$, sensitive to both Fourier amplitudes and phases, 2LPT performs a couple of per cent better than \muscle\ at $k\sim0.2$\hmpc. Thus, even though \muscle's detection of large-scale collapses somehow manages to treat the large-scale tidal field with some accuracy, 2LPT still seems to be the best available method for small fluctuations, e.g.\ to generate high-redshift initial conditions.

However, \muscle\ outperforms both perturbative schemes on small scales. Interpolating in scale between 2LPT and \muscle\ (`\muscle+2LPT') gives the highest $R(k)$ at all scales. Interpolating between 2LPT and \SC\ (labeled `\alpt') gives results only slightly worse than interpolating with \muscle. Note that due to cosmic variance, small differences in $R(k)$ may not persist for all realizations (see KH13 for error estimates.) However, note that the $R(k)$ curves may (co)vary together up and down more than independently, so small differences may be more statistically significant than the error bars suggest. The superiority of all Lagrangian schemes is obvious from the low `Eulerian linear theory' curve \citep{TassevZaldarriaga2012}, essentially the propagator \citep{CrocceScoccimarro2006}.

The bottom panel of Fig.\ \ref{fig:crosscorr} shows the ratio of the nonlinear power spectrum to those in the various approximations. The power spectra in the $\psi$-based schemes differd only slightly, but the approximation with power closest to full $N$-body is `\muscle+2LPT.' Notably, the large-scale power deficiency in \SC\ is fixed in \muscle. In both measurements, while \muscle\ performs well, there is still much distance left to go to `Perfect $\psi$,' which would be the result if the final $\psi$ could be perfectly predicted.

\begin{figure}
   \begin{center}
    \includegraphics[width=\columnwidth]{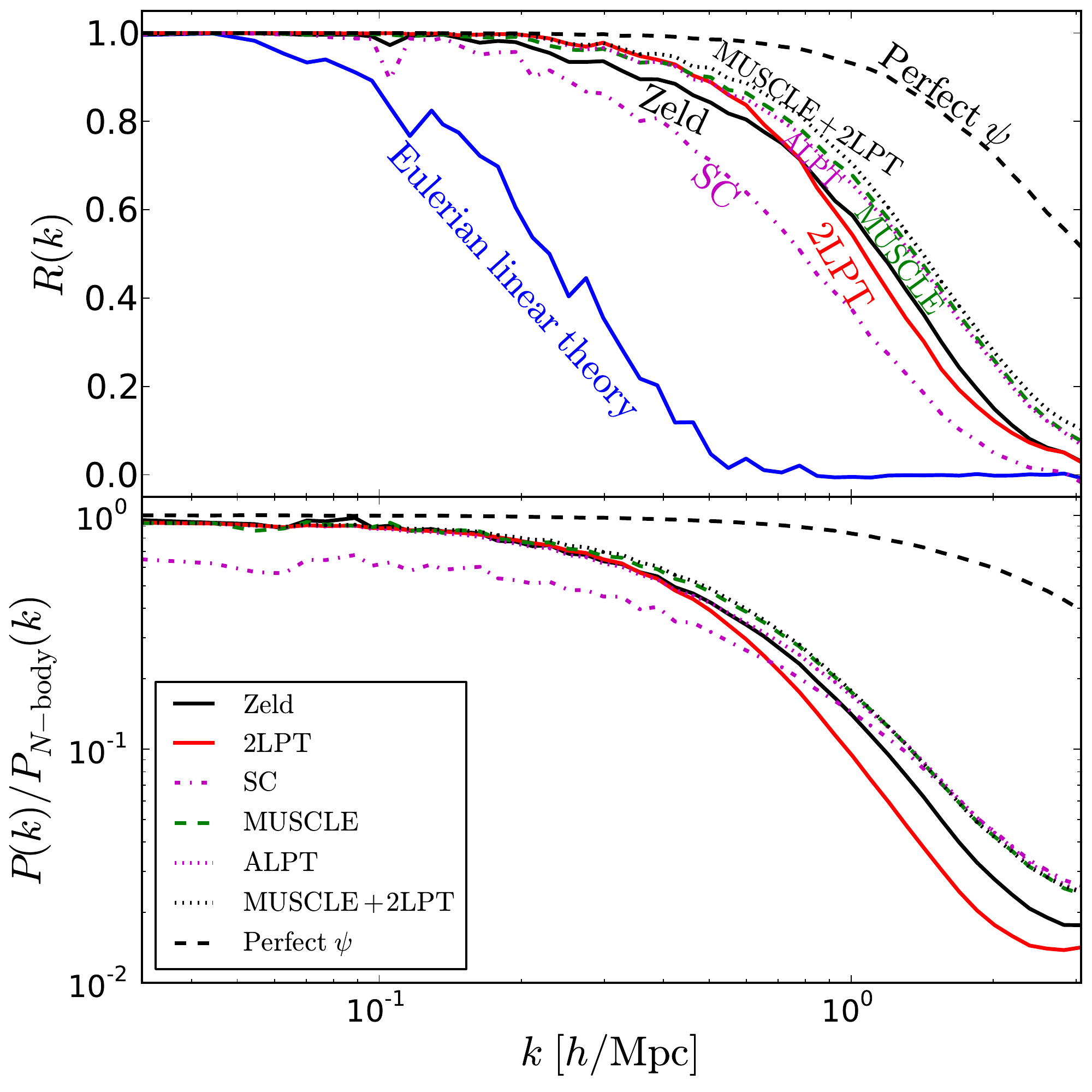}
  \end{center}  
  \caption{{\it Top:} Fourier-space cross-correlations of the $N$-body density fields with the various approximations. {\it Bottom}: Ratios of power spectra of the various approximations to that of the full $N$-body density field.}
  \label{fig:crosscorr}
\end{figure}

\begin{figure}
   \begin{center}
    \includegraphics[width=\columnwidth]{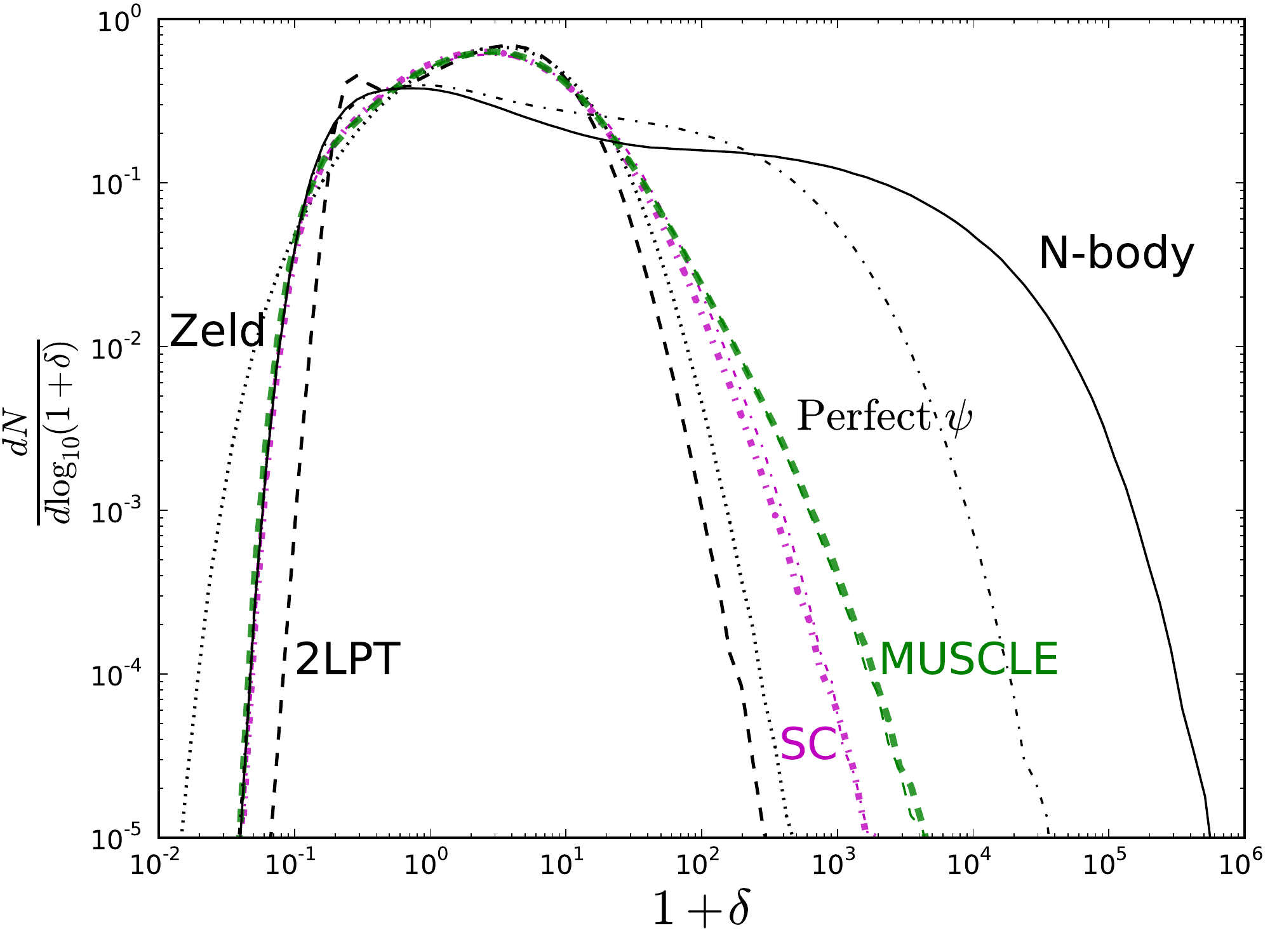}
  \end{center}  
  \caption{Mass-weighted density PDF's of the various realizations, measured using a Voronoi tessellation. The thick magenta dashdotted curve shows the PDF using raw \SC, and the thick dashed and curve shows the PDF using \muscle. The nearly overlapping thin curves of the same colors show PDF's in \alpt\ and `2LPT+\muscle,' i.e.\ mixing in 2LPT on large scales.}
  \label{fig:musclepdfs}
\end{figure}

\muscle\ also gets the 1-point PDF of the field far more accurately than the perturbative approaches. Fig.\ \ref{fig:musclepdfs} shows mass-weighted histograms of densities computed using the Voronoi Tessellation Field Estimator \citep[e.g.][]{svdw}, with each particle contributing once to the histogram. The agreement on the low-density tail between the $N$-body realization and the \SC\ and \muscle\ realizations is remarkable. Note even the unphysical low-density peak in 2LPT. At high densities, \SC\ and \muscle\ also perform the most accurately of the approximations, although they still fall well short of $N$-body. The \muscle\ PDF is the closest of the approximations, evidently since it captures additional large-scale collapses compared to \SC. But even `Perfect $\psi$' falls somewhat short. Adding large-scale 2LPT modes to \SC\ and \muscle\ gives almost no change in these PDF's. 

\begin{figure}
   \begin{center}
    \includegraphics[width=\columnwidth]{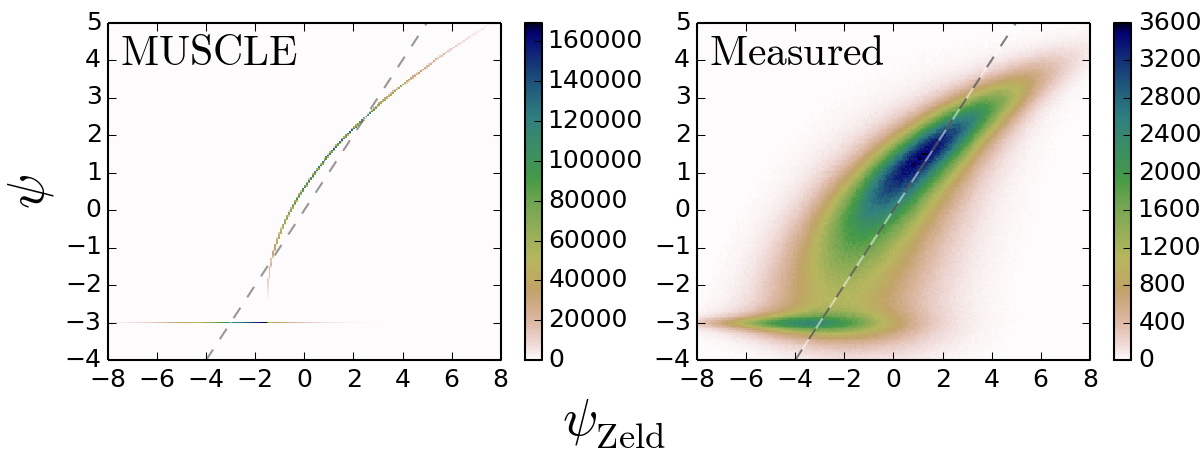}
  \end{center}  
  \caption{A comparison of $\psi=\divqpsi$ as predicted from the initial conditions using \muscle, to that measured in an $N$-body simulation. $\psi_{\rm Zeld}$, on the $x$-axis of both plots, is the linear-theory (Zel'dovich) prediction. The colorbar numbers indicate the number of particles (out of 256$^3$) in that two-dimensional bin. The white-and-black line indicates where the Zel'dovich prediction would lie.}
  \label{fig:psiscatters}
\end{figure}

Fig.\ \ref{fig:psiscatters} shows 2D histograms of $\psi$, both as predicted using \muscle, and as measured at $z=0$ in the simulation, versus its linear-theory prediction from the initial conditions, $\psi_{\rm Zeld}$. The relationship is simple in \muscle: if $\psi_{\rm Zeld}<-1.5$, $\psi_{\rm musc}=-3$. But there are also particles with $\psi_{\rm Zeld}>-1.5$ that get mapped to $-3$; for these, $\psi_{\rm Zeld}$ dips below $-1.5$ when smoothing on some scale. There are two ways used in this paper to measure the divergence $\psi$ from the final conditions: with an FFT; and a `difference method,' in which $\psi$ is measured at a particle by differencing particle positions initially on either side of it (this has a resolution twice as coarse as the FFT method).  The results shown here and in the next figure use the difference method, while the `Perfect $\psi$' results used elsewhere use the FFT method (which does not give a clear locus of points at $\psi=-3$, as seen here).

Another interesting measurement is the distribution of $\psi$ in the $N$-body simulation for different morphologies of particles: void, wall, filament and halo. To classify particles, we use the \origami\ algorithm \citep{FalckEtal2012}, which assigns the morphology of a particle according to the number of orthogonal axes by which any other particle crosses it (void=0, wall=1, filament=2, and halo =3), comparing the initial and final conditions.

\begin{figure}
   \begin{center}
    \includegraphics[width=\columnwidth]{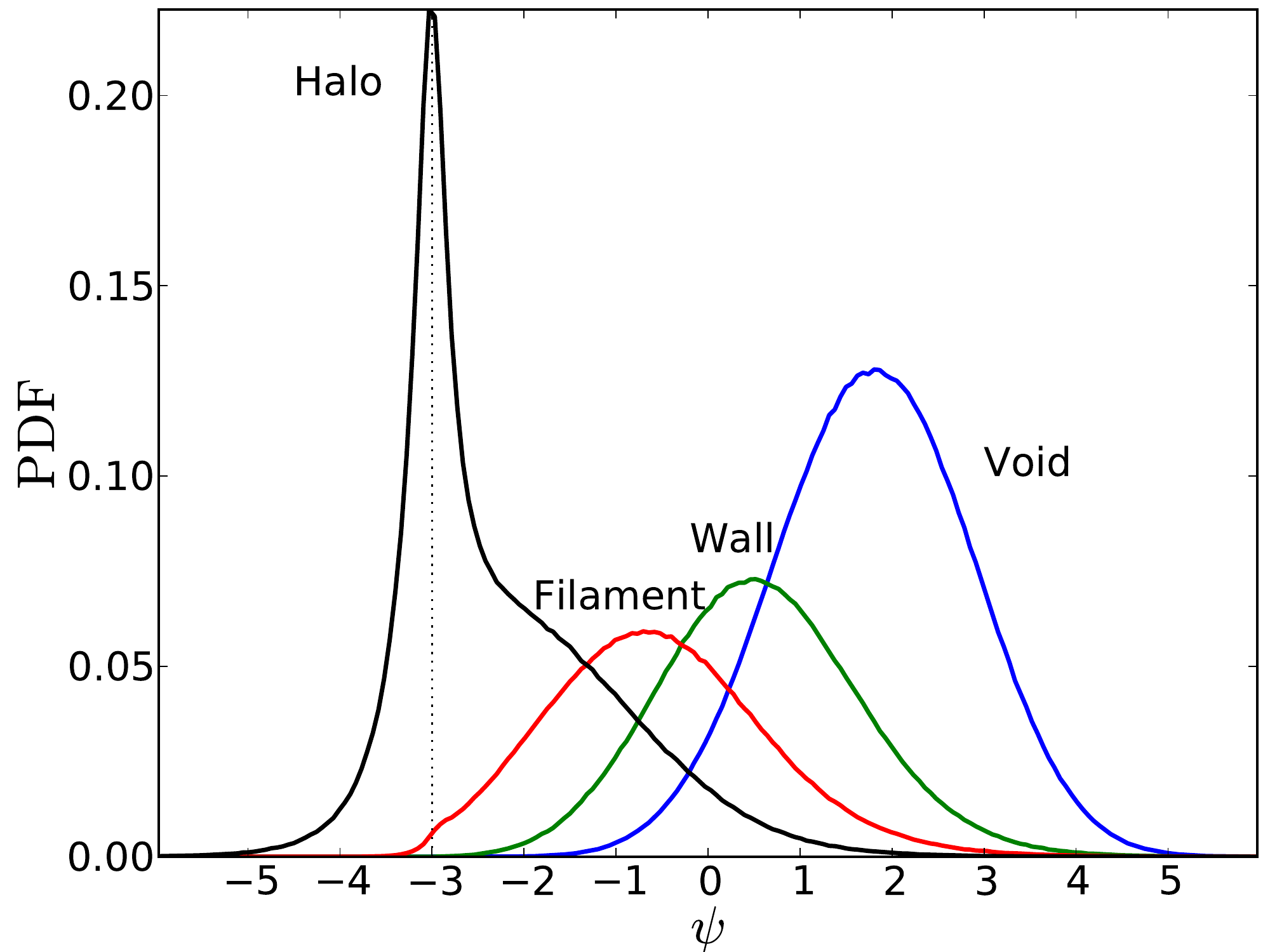}
  \end{center}  
  \caption{Stretch parameters $\psi=\divqpsi$ measured in an $N$-body simulation of interparticle spacing $\sim 1$\hmpc, separated by morphological type. The distribution for each type looks rather Gaussian, except for `Halo;' see discussion in text.}
  \label{fig:depend}
\end{figure}

Fig.\ \ref{fig:depend} shows these PDF's for the various morphologies. Except for `Halo,' the distributions of $\psi$ look quite Gaussian. One part of the `Halo' distribution is a sharp spike at $\psi=-3$, characterizing a Lagrangian patch that precisely contracts to an Eulerian point. There is another component that looks somewhat Gaussian, as well; it would be interesting to differentiate these populations physically. One possibility is that the high-$\psi$ tail consists of particles on their first infall. But also, the `Filament' population has a small bump at $\psi=-3$, suggesting a small amount of contamination between the two morphologies as detected by \origami.

The positions of the other curves are also informative. From void to halo, the mean $\psi$'s of each morphology are $\psi=(1.80, 0.58, -0.54, -2.25)$. If wall and filament volume elements typically collapsed along one and two axes, their non-collapsing dimensions staying fixed in comoving coordinates, their means would be at $\psi=-1$ and $\psi=-2$. Their higher $\psi$ values indicate that their non-collapsing dimensions tend to stretch substantially.

\section{Discussion and Conclusion}

Here we have presented a conceptually simple, non-perturbative Lagrangian scheme that produces fast $N$-body realizations, essentially as fast as producing initial conditions of a simulation, and more accurately than other such schemes. It applies a spherical-collapse criterion to the Lagrangian divergence of the displacement field, $\psi$, on the pixel scale of the initial conditions, and on larger scales as well. Approximate $N$-body schemes have proven useful to produce mock galaxy catalogs \citep{KitauraEtal2014patchy}. Some past and upcoming surveys include emission-line galaxies that do not necessarily occupy the largest dark-matter haloes, such as WiggleZ \citep{DrinkwaterEtal2010} and eBOSS \citep{ComparatEtal2013}.  For these, fast approximations which accurately resolve the cosmic web to small scales will be increasingly important, for example to produce mock galaxy catalogs with a biasing prescription \citep[e.g.][]{KitauraEtal2014patchy,AtaEtal2015}.

There is still room for improvement in such a $\psi$-based scheme; if it were possible to perfectly predict $\psi$, realization accuracies would increase substantially. One possible way forward is to predict the final morphologies (void, wall, filament, halo) of particles, and assign $\psi$ to each differently, according to their values in Fig.\ \ref{fig:depend}. Note that although \SC\ and \muscle\ collapse haloes rather well, filaments do not visibly tighten substantially compared to the ZA, at the resolution presented here (at higher resolution, they would tighten some filaments, since \muscle\ limits overcrossing).  However, a separate prescription for each morphological type would sacrifice conceptual simplicity, and likely introduce some ad-hoc parameters.  Another idea is to remap the approximate realization's matter-density PDF to that of a full $N$-body simulation \citep{LeclercqEtal2013}; note that this step would be bypassed if generating a mock galaxy catalog.  Also, we found that interpolating between 2LPT and \muscle\ as in \alpt\ can slightly improve accuracy over \muscle, at the expense of slightly adding to the conceptual complexity and the run time.

A Python package that generates all $\psi$-based particle realizations discussed here, including an \alpt\ interpolation in scale, is available at \url{http://skysrv.pha.jhu.edu/~neyrinck/muscle.html}. It interfaces with \camb\ via the \cosmopy\ package, at \url{http://www.ifa.hawaii.edu/cosmopy/}.

\section*{Acknowledgments}
I thank Francisco-Shu Kitaura for helpful conversations, and Miguel Arag\'{o}n-Calvo for use of the simulation analyzed here.  I am grateful for financial support from a grant in Data-Intensive Science from the Gordon and Betty Moore and Alfred P. Sloan Foundations.

\bibliographystyle{mnras}
\bibliography{refs}

\end{document}